\documentclass{article}
\usepackage{hello}

\usepackage[margin=1in]{geometry}

\usepackage{graphicx}
\usepackage{booktabs}
\usepackage{array}
\usepackage{amsmath}
\usepackage{bbm}
\usepackage{amssymb}
\usepackage{amsfonts}
\usepackage{multirow}
\usepackage{verbatim}
\usepackage{graphicx}
\usepackage{caption}
\usepackage{url}
\usepackage{longtable}
\usepackage{supertabular}
\usepackage{float}
\usepackage{enumitem}
\usepackage{tablefootnote}
\usepackage[round,semicolon]{natbib}
\usepackage{xcolor}
\usepackage{colortbl}
\usepackage{xspace}
\usepackage{textcomp}
\usepackage{makecell}
\usepackage{multirow}
\usepackage{lscape} 
\usepackage{siunitx}
\usepackage{tabularx}

\usepackage{amssymb}% http://ctan.org/pkg/amssymb
\usepackage{pifont}% http://ctan.org/pkg/pifont
\usepackage{scrextend}

\usepackage{array}
\usepackage{tgpagella}
\usepackage{latexsym}
\usepackage[T1]{fontenc}
\usepackage[utf8]{inputenc}
\usepackage{microtype}
\definecolor{mydarkblue}{rgb}{0,0.08,0.45}
\usepackage[colorlinks,citecolor=mydarkblue,urlcolor=mydarkblue,linkcolor=mydarkblue]{hyperref}
\usepackage{url}            % simple URL typesetting
\usepackage{nicefrac}       % compact symbols for 1/2, etc.
\usepackage{changepage}
\usepackage{xargs}          % Use more than one optional parameter in a new commands
\usepackage{wrapfig,lipsum,booktabs}
\usepackage{longtable}
\usepackage{caption}
\usepackage{subcaption}
\usepackage{endnotes}
\usepackage{tablefootnote} % for table footnotes

\usepackage{pgfplots}
\usetikzlibrary{pgfplots.groupplots}
\pgfplotsset{compat=1.3}
\usepackage{tikz}
\usetikzlibrary{patterns}

\usepackage[most]{tcolorbox}

\usepackage[capitalize,noabbrev]{cleveref}
\crefname{section}{Section}{\S\S}
\Crefname{section}{Section}{\S\S}
\crefname{table}{Table}{Tables}
\crefname{figure}{Figure}{Figures}
\crefname{algorithm}{Algorithm}{}
\crefname{equation}{eq.}{}
\crefname{appendix}{Appendix}{}
\crefformat{section}{Section #2#1#3}
\usepackage{multicol}
\usepackage{multirow}
\usepackage{bm}

\usepackage{tcolorbox}
\usepackage{titlesec}
\titleformat*{\section}{\large\bfseries}

\definecolor{battleshipgrey}{rgb}{0.3, 0.3, 0.3}
\definecolor{brilliantrose}{rgb}{1.0, 0.33, 0.64}
\definecolor{americanrose}{rgb}{1.0, 0.01, 0.24}
\definecolor{jweigreen}{rgb}{0,0.45,0.24}
\definecolor{bluegray}{rgb}{0.1, 0.1, 0.4}
\definecolor{ao(english)}{rgb}{0.0, 0.5, 0.0}
\definecolor{blanchedalmond}{rgb}{1.0, 0.92, 0.8}
\definecolor{atomictangerine}{rgb}{1.0, 0.6, 0.4}
\definecolor{chocolate(web)}{rgb}{0.82, 0.41, 0.12}
\definecolor{bananayellow}{rgb}{1.0, 0.88, 0.21}
\definecolor{goldenbrown}{rgb}{0.6, 0.4, 0.08}
\definecolor{aliceblue}{rgb}{0.94, 0.97, 1.0}
\definecolor{beige}{rgb}{0.96, 0.96, 0.86}
\definecolor{babyblue}{rgb}{0.54, 0.81, 0.94}
\definecolor{camel}{rgb}{0.76, 0.6, 0.42}
\definecolor{cinnamon}{rgb}{0.82, 0.41, 0.12}
\definecolor{deepskyblue}{rgb}{0.0, 0.75, 1.0}
\definecolor{frenchblue}{rgb}{0.0, 0.45, 0.73}
\definecolor{classicrose}{rgb}{0.98, 0.8, 0.91}
\definecolor{frenchrose}{rgb}{0.96, 0.29, 0.54}
\definecolor{frenchlilac}{rgb}{0.53, 0.38, 0.56}
\definecolor{frenchbeige}{rgb}{0.65, 0.48, 0.36}
\definecolor{verylightgreen}{RGB}{240, 255, 235}
\definecolor{verylightred}{RGB}{255, 235, 235}
\definecolor{verylightyellow}{RGB}{255, 254, 235}
\definecolor{dt}{gray}{0.7}

\definecolor{forestgreen}{HTML}{2e7d43}
\definecolor{color1}{HTML}{FF9999}
\definecolor{color2}{HTML}{FF6666}
\definecolor{color3}{HTML}{FF3333}
\definecolor{color4}{HTML}{E60000}
\definecolor{color5}{HTML}{B30000}
\definecolor{color6}{HTML}{8CD98C}
\definecolor{color7}{HTML}{53c653}
\definecolor{color8}{HTML}{39ac39}
\definecolor{color9}{HTML}{2d862d}
\definecolor{color10}{HTML}{206020}
\definecolor{color11}{HTML}{cca300}

\usepackage{minitoc}

\title{
\textbf{
Qwen-Audio: Advancing Universal Audio Understanding via Unified Large-Scale Audio-Language Models}
}
\author{
\large{}
Yunfei Chu$^*$ \hspace{6mm} Jin Xu$^*$ \hspace{6mm} Xiaohuan Zhou$^*$ \hspace{6mm} Qian Yang\\
Shiliang Zhang \hspace{6mm} Zhijie Yan \hspace{6mm} Chang Zhou$^{\dag}$ \hspace{6mm} Jingren Zhou
\\
\large{}
Alibaba Group
\\
\\
\small{}
Code \& Demo \& Models: \ \ \url{https://github.com/QwenLM/Qwen-Audio}
}

\date{}

\begin{document}

\doparttoc % Tell to minitoc to generate a toc for the parts
\faketableofcontents % Run a fake tableofcontents command for the partocs

\maketitle

\begin{abstract}
\noindent
Recently, instruction-following audio-language models have received broad attention for audio interaction with humans.
However, the absence of pre-trained audio models capable of handling diverse audio types and tasks has hindered progress in this field. Consequently, most existing works have only been able to support a limited range of interaction capabilities. In this paper, we develop the Qwen-Audio model and address this limitation by scaling up audio-language pre-training to cover over 30 tasks and various audio types, such as human speech, natural sounds, music, and songs, to facilitate universal audio understanding abilities. 
However, directly co-training all tasks and datasets can lead to interference issues, as the textual labels associated with different datasets exhibit considerable variations due to differences in task focus, language, granularity of annotation, and text structure. 
To overcome the one-to-many interference, we carefully design a multi-task training framework by conditioning on a sequence of hierarchical tags to the decoder for encouraging knowledge sharing and avoiding interference through shared and specified tags respectively. Remarkably, Qwen-Audio achieves impressive performance across diverse benchmark tasks without requiring any task-specific fine-tuning, surpassing its counterparts. Building upon the capabilities of Qwen-Audio, we further develop Qwen-Audio-Chat, which allows for input from various audios and text inputs, enabling multi-turn dialogues and supporting various audio-central scenarios.

\end{abstract}

{\let\thefootnote\relax\footnotetext{$^*$Equal contribution, $^\dag$Corresponding author}}

\begin{figure*}[t!]
\centering
\includegraphics[width=12cm]{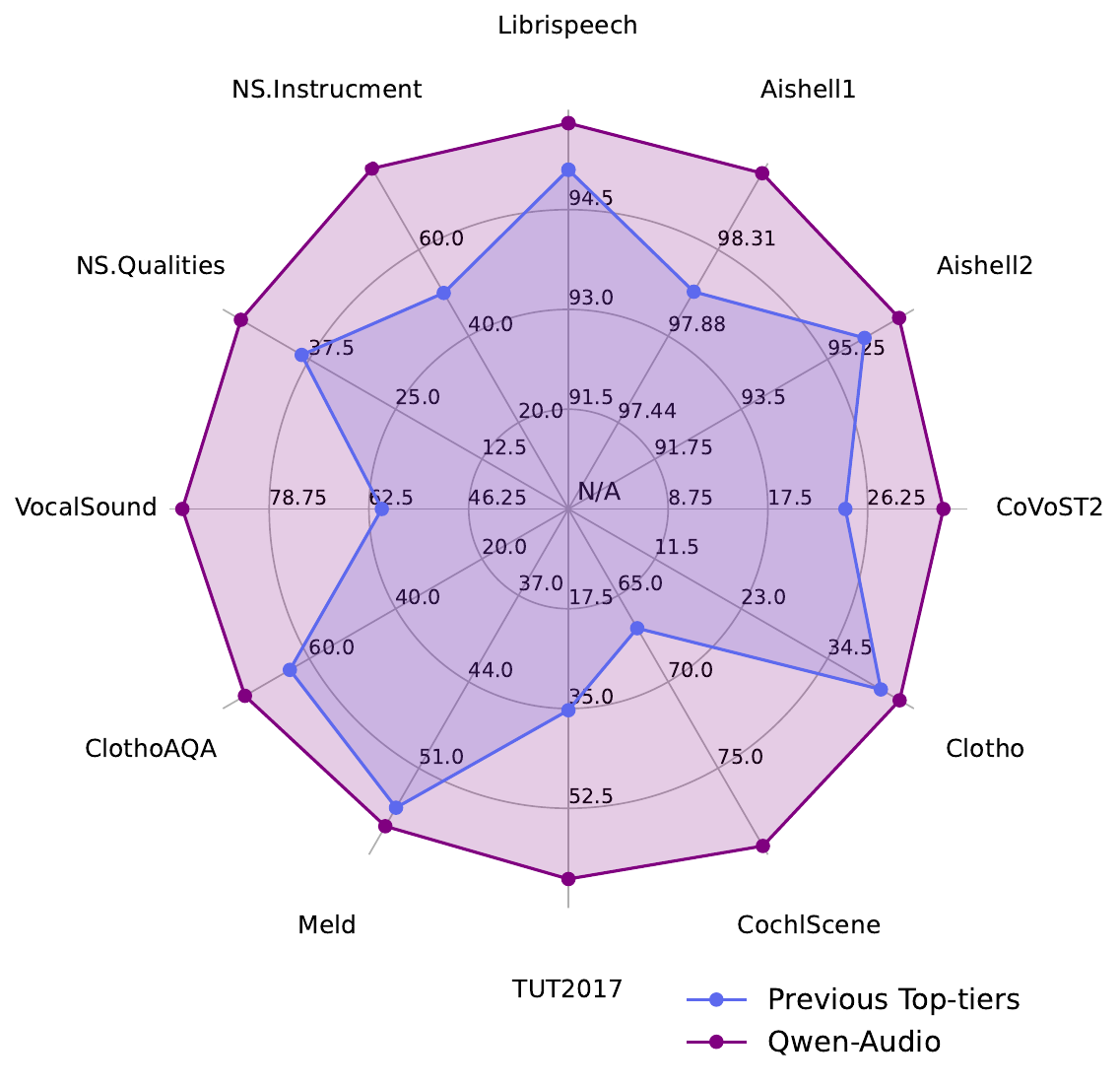}
\caption{Performance of Qwen-Audio and previous top-tiers from multi-task audio-text learning models such as SpeechT5~\citep{ao2021speecht5}, SpeechNet~\citep{chen2021speechnet}, SpeechLLaMA~\citep{speechllama}, SALMONN~\citep{anonymous2023salmonn} and Pengi~\citep{Pengi}. We demonstrate the test set results across the 12 datasets covering Automatic Speech Recognition~(ASR), Speech-to-Text Translation~(S2TT), Automatic Audio Captioning~(AAC), Acoustic Scene Classification~(ASC), Speech Emotion Recognition~(SER), Audio Question and Answering~(AQA), Vocal Sound Classification~(VSC), and Music Note Analysis~(MNA). The results of ASR datasets, such as Librispeech, Aishell1 and Aishell2 refer to 1 - WER\%. The results of CoVoST2 is the average BLEU score of seven translation directions (en-de, de-en, en-zh, zh-en, es-en, fr-en and it-en). Qwen-Audio achieves remarkable performance without requiring any task-specific fine-tuning, surpassing its counterparts.}
\label{img:radar}
\end{figure*}

\begin{figure*}[t!]
\centering
\includegraphics[width=\textwidth]{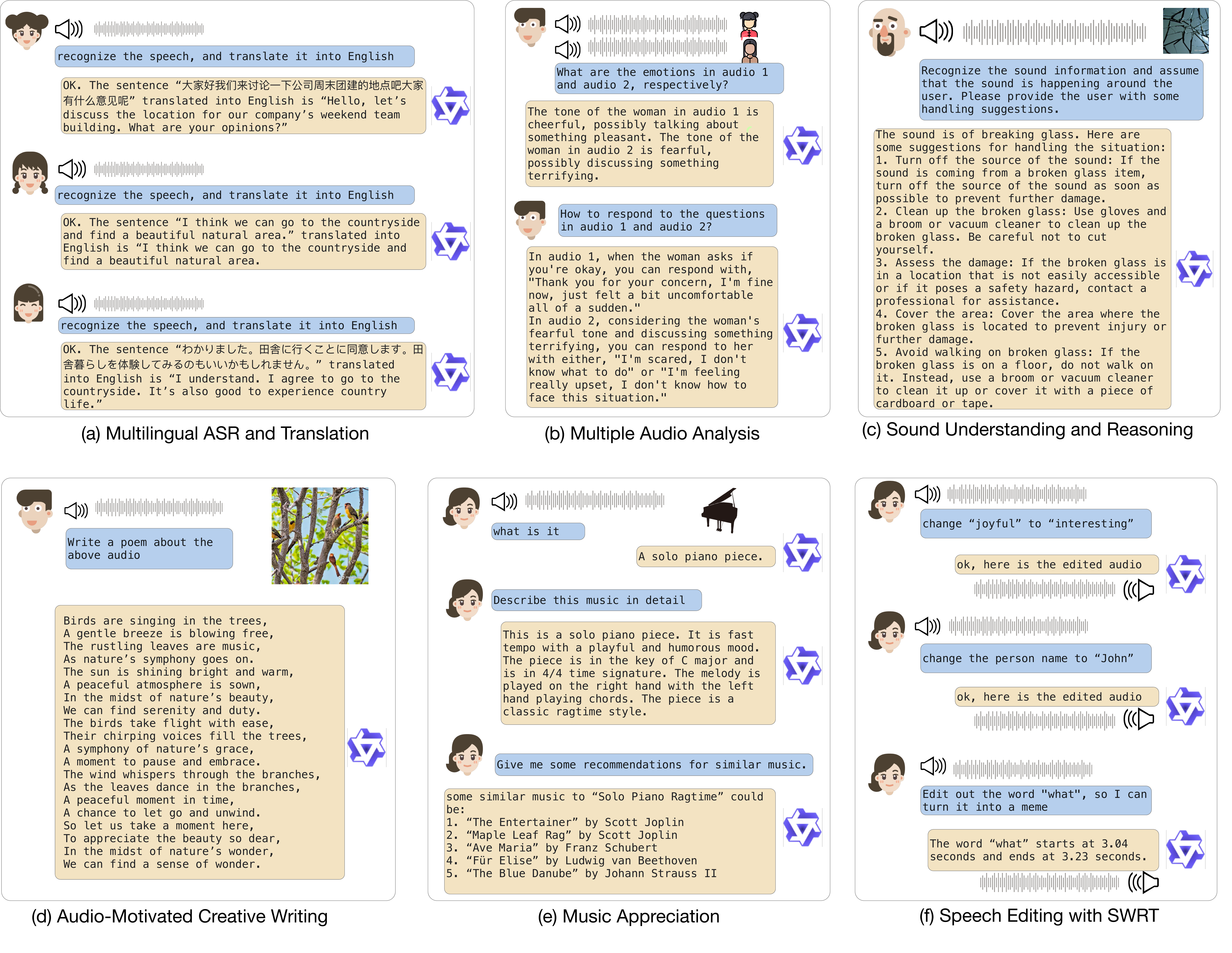}
\caption{Examples of Qwen-Audio showcasing its proficiency in perceiving and comprehending various types of audio. Qwen-Audio supports multiple-audio analysis, sound understanding and reasoning, music appreciation, and tool usage for speech editing. Demos are available at \url{https://qwen-audio.github.io/Qwen-Audio/}.}
\label{img:demo}
\end{figure*}

\section{Introduction}
Large language models (LLMs)~\citep{gpt3, chatgpt, gpt4, palm, palm2, llama, llama2, qwen7b} have greatly propelled advancements in the field of general artificial intelligence (AGI) due to their strong knowledge retention, complex reasoning and problem-solving capabilities. However, language models lack the capability to perceive non-textual modalities like images and audio in the same manner as humans do. Speech, as an important modality, provides diverse and complex signals beyond texts such as emotions, tones, and intentions in human voice, train whistle, clock chime and thunder in natural sounds, and melody in music. Enabling LLMs to perceive and comprehend rich audio signals for audio interaction has received broad attention~\citep{AudioGPT, HuggingGPT, wang2023blsp, Macaw-LLM, NextGPT, LTU, wang2023slm, shu2023llasm}.

Prior works for instruction following mainly inherit the capabilities from large (multimodal) LLMs and adopt light-weight supervised fine-tuning to activate the abilities of the model to align with user intent~\citep{ouyang2022intructgpt, wang2023blsp, LTU}. However, most works have been constrained in terms of audio interaction capabilities due to the lack of  pre-trained audio-language models that can handle diverse audio types and tasks. Existing representative audio-language multi-tasks language models, such as SpeechNet~\citep{chen2021speechnet}, SpeechT5~\citep{ao2021speecht5}, VIOLA~\citep{VIOLA}, Whisper~\citep{Whisper} and Pengi~\citep{Pengi} are limited to processing specific audio types, such as human speech or natural sounds.

To promote the growth and development of the audio-text multimodal community, we introduce Qwen-Audio, a large-scale audio-language model. Qwen-Audio is a multi-task language model conditioning on audio and text inputs, that extends the Qwen-7B~\citep{bai2023qwen} language model to effectively perceive audio signals by the connection of a single audio encoder. Different from previous works that primarily cater to a single type of audio such as human speech, or focus on specific tasks like speech recognition and captioning, or limit models to a single language~\citep{wang2023blsp, Macaw-LLM, NextGPT, LTU, shu2023llasm}, we scale up the training to dozens of datasets covering over 30 tasks, eight languages and various types of audio for advancing universal audio understanding abilities. A significant challenge of multi-task and multi-dataset co-training arises from the considerable variation in textual labels associated with different datasets. This variation stems from differences in task objectives, languages, annotation granularity, and text structure (structured or unstructured). To address this one-to-many challenge, we have carefully designed a multi-task training framework that conditions the decoder on a sequence of hierarchical tags. This design encourages knowledge sharing and helps mitigate interference through shared and specified tags, respectively. Furthermore, we incorporate the speech recognition with the word-level time-stamp prediction (SRWT) task for training, which is usually ignored in previous multi-task learning research. We find the task not only improves the grounding and grounding-based QA tasks beyond speech signals such as sound and music, but also improves the performance of ASR. As shown in Figure~\ref{img:radar}, extensive evaluation demonstrates that Qwen-Audio, without any task-specific fine-tuning, outperforms previous multi-task training models across a diverse range of tasks. A notable achievement of Qwen-Audio is its state-of-the-art performance on the test set of Aishell1, cochlscene, ClothoAQA, and VocalSound. Leveraging the capabilities of Qwen-Audio, we introduce Qwen-Audio-Chat via supervised instruction fine-tuning, which facilitates flexible input from both audio and text modalities in multi-turn dialogues, enabling effective interaction following human instructions. The contribution of the paper is summarized below:

\begin{itemize}
%yq part
\item We introduce Qwen-Audio, a fundamental multi-task audio-language model that supports various tasks, languages, and audio types, serving as a universal audio understanding model. Building upon Qwen-Audio, we develop Qwen-Audio-Chat through instruction fine-tuning, enabling multi-turn dialogues and supporting diverse audio-oriented scenarios. Both Qwen-Audio and Qwen-Audio-Chat models are open-source, promoting the growth and development of the audio-text multimodal community. 
\item To scale up audio-language pre-training, we address the challenge of variation in textual labels associated with different datasets by proposing a multi-task training framework, enabling knowledge sharing and avoiding one-to-many interference. Our model incorporates more than 30 tasks and extensive experiments show the model achieves strong performance.
\item To promote audio-language pre-training, we demonstrate that incorporating the SRWT task, which is often overlooked in the audio multimodal research community, improves grounding and grounding-based question answering tasks beyond speech signals, as well as ASR performance.
\item Experimental results show that Qwen-Audio achieves impressive performance across diverse benchmark tasks without requiring any task-specific fine-tuning, surpassing its counterparts. Specifically, Qwen-Audio achieves state-of-the-art results on the test set of Aishell1, cochlscene, ClothoAQA, and VocalSound. 

\end{itemize}

\section{Related Work}
\label{sec:related_work}
\paragraph{Multi-task Audio-Text Learning}
The goal of multi-task training is to transfer knowledge between different tasks with unified model architectures and data format~\citep{raffel2020exploring,ao2021speecht5,chen2021speechnet}. In audio processing domains, it is challenging to unify all audio processing tasks since there are various audio signals, such as human speech, natural sounds, music, and songs, and their labeling format differs a lot. SpeechNet~\citep{chen2021speechnet} and SpeechT5~\citep{ao2021speecht5} treat human speech tasks into a speech/text input and speech/text output format, and leverage a shared encoder-decoder framework for pretraining. Many works~\citep{VIOLA, maiti2023voxtlm, AudioPALM, SpeechX, LMWithVoice} unify data format and tasks by directly feeding speech representation~\citep{LMWithVoice} or encoding continuous speech signals as discrete codes~\citep{Encodec, SoundStream,USM}, and treating different human speech tasks as conditional generative tasks. For training, they directly adopt a decoder-only Transformer model~\citep{Transformer}. VoiceBox~\citep{VoiceBox} employs a non-autoregressive continuous normalizing flow model for human speech synthesis and speech editing tasks. 
Whisper~\citep{Whisper} proposes a template for multi-task training, considering the granularity of dataset annotations (with or without sentence-level timestamps) and task types (human speech recognition and translation) for unified training. Previous works mostly focus only on human speech processing tasks such as speech recognition and translation and ignore other types of audio such as natural sounds and music. Pengi~\citep{Pengi} focuses on natural sound understanding tasks and treats these tasks as text generation tasks. Specifically, Pengi unifies data format with text templates and then trains all tasks within a Transformer decoder model. In this work, Qwen-Audio integrates diverse audio types, such as human speech, natural sounds, music, and songs, and facilitates co-training on datasets sourced from heterogeneous data and featured disparate labeling granularities. This is achieved through introducing a unified learning framework. Upon completion of the co-training process, the model demonstrates comprehensive capabilities in speech perception, comprehension, and recognition tasks, eliminating the need for additional task-specific architectural extensions.

\paragraph{Interact with LLMs through Multiple Modality}
Recently, large language models such as ChatGPT~\citep{chatgpt} have demonstrated impressive capabilities for knowledge retention, reasoning, and coding followed by human instructions. To extend to application scope of LLMs beyond pure text tasks, many LLM-based multimodal models have been developed. For visual modality, GPT4~\citep{gpt4}, Flamingo~\citep{alayrac2022flamingo}, Kosmos~\citep{kosmos2}, BLIP~\citep{blip}, Shikra~\citep{shikra}, Emu~\citep{EMU} and Qwen-VL~\citep{qwenvl} have proposed different integration method to enable image understanding or generation capabilities for LLMs.

For the audio modality, there have been attempts to utilize well-trained audio foundation models as tools, such as AudioGPT~\citep{AudioGPT} and HuggingGPT~\citep{HuggingGPT}, while leveraging LLMs as a versatile interface. These endeavors involve instructing LLMs to generate commands for controlling external tools or transcribing human speech to text before inputting into the LLMs. However, these approaches lack the inclusion of crucial information like prosody and sentiment in human speech, and in certain cases, they fail to convert non-textual audio, such as natural sound. Consequently, the transfer of knowledge from LLMs to the speech modality encounters obstacles, and the LLMs lack the necessary capabilities to perceive and comprehend audio signals. Recent efforts explore training end-to-end audio-text LLMs for direct speech interaction. SpeechGPT~\citep{speechgpt} first converts human speech into discrete HuBERT tokens~\citep{HuBERT}, and then designs a three-stage training pipeline on paired speech data, speech instruction data and chain-of-modality instruction data accordingly. BLSP~\citep{wang2023blsp} aligns representation by requiring the LLM to generate the same text continuation given the human speech and corresponding transcripts.
LLaSM~\citep{shu2023llasm} creates large speech instruction datasets by generating speech questions using Microsoft TTS API, and then conducts training to enable end-to-end interaction between human speech and text.

LTU~\citep{LTU} creates a 5M audio QA dataset, and conducts supervised finetuning (SFT) on the audio module and LoRA adapters~\citep{hu2021lora} of LLaMA~\citep{touvron2023llama} to enhance the alignment between sound perception and reasoning. SALMMON~\citep{anonymous2023salmonn} utilizes both a text encoder and a speech encoder to extract the representation from different kinds of audio and text input, and then connects the inputs to a well-train LLM with Q-former~\citep{blip2} style attention to generate response.
In this work, Qwen-Audio aims at training a unified audio-text multi-task multilingual LLMs capable of perceiving and understanding audio inputs while preserving the textual conversational abilities. Qwen-Audio employs a single encoder for all audios, and bridges the gap of audio and text modality by large-scale end-to-end training to support various tasks such as natural sound detection, human speech recognition and grounding, and audio captions tasks. The resulting model demonstrates superior performance than previous works across a diverse style of tasks.

\section{Methodology}
\label{method}
This section provides details of Qwen-Audio and Qwen-Audio-Chat, which are designed for universal audio understanding and flexible interaction based on human instructions. The model structure of Qwen-Audio and Qwen-Audio-Chat is first presented in Section~\ref{exp:model_arch}. The training process of our models consists of two stages: multitask pretraining and supervised fine-tuning. We describe the training of Qwen-Audio via multitask learning in Section~\ref{exp:multitask_train}. Then, we describe Qwen-Audio-Chat with supervised fine-tuning in Section~\ref{exp:sft} , which enables flexible human interaction.

\begin{figure*}[t!]
\centering
\includegraphics[width=15cm]{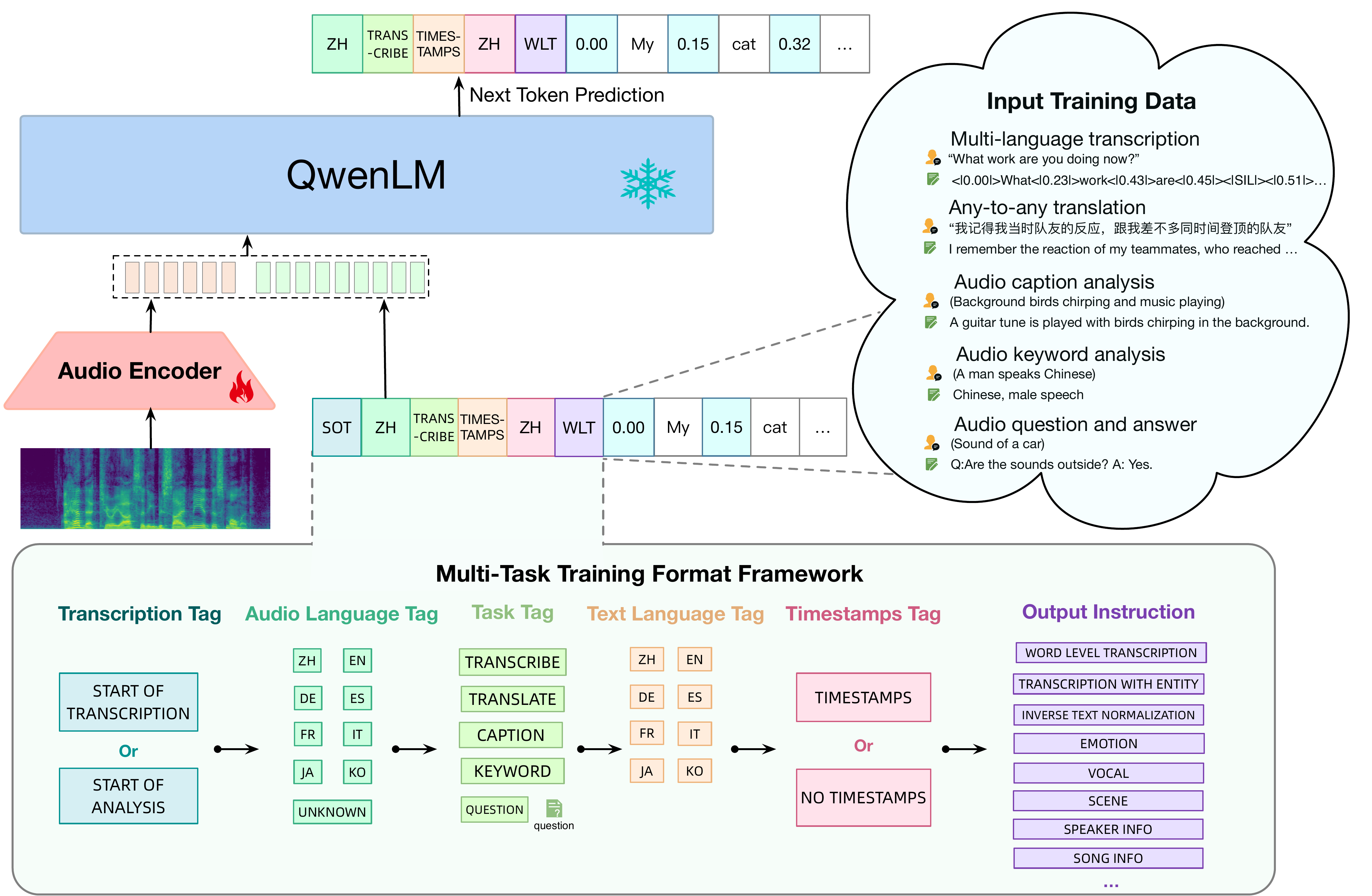}
   \caption{The overview of Qwen-Audio architecture and multitask-pretraining.}
\label{img:format}
\end{figure*}

\subsection{Model Architecture}\label{exp:model_arch}
The architecture of Qwen-Audio models is depicted in Figure~\ref{img:format}. Qwen-Audio contains an audio encoder and a large language model. Given the paired data $(\bm{a}, \bm{x})$, where the $\bm{a}$ and $\bm{x}$ denote the audio sequences and text sequences, the training objective is to maximize the next text token probability as
\begin{equation}
    \mathcal{P}_{\theta}(x_t|\bm{x}_{<t}, \text{Encoder}_{\phi}(\bm{a})),
\end{equation}
conditioning on audio representations and previous text sequences $\bm{x}_{<t}$, where $\theta$ and $\phi$ denote the trainable parameters of the LLM and audio encoder respectively. 

\paragraph{Audio Encoder}
Qwen-Audio employs a single audio encoder to process various types of audio. The initialization of the audio encoder is based on the Whisper-large-v2 model~\citep{Whisper}, which is a 32-layer Transformer model that includes two convolution down-sampling layers as a stem. The audio encoder is composed of 640M parameters. Although Whisper is supervised trained for speech recognition and translation, its encoded representation still contains rich information, such as background noise~\citep{whisperAT}, and can even be used for recovering the original speech~\citep{SpeechTokenizer}.
To preprocess the audio data, Whisper resamples it to a frequency of 16kHz and converts the raw waveform into 80-channel mel-spectrogram using a window size of 25ms and a hop size of 10ms. Additionally, a pooling layer with a stride of two is incorporated to reduce the length of the audio representation. As a result, each frame of the encoder output approximately corresponds to a 40ms segment of the original audio signal. SpecAugment~\citep{specaug} is applied at the training time as data augmentation.

\paragraph{Large Language Model}
Qwen-Audio incorporates a large language model as its foundational component. The model is initialized using pre-trained weights derived from Qwen-7B~\citep{bai2023qwen}. Qwen-7B is a 32-layer Transformer decoder model with a hidden size of 4096, encompassing a total of 7.7B parameters.

\subsection{Multitask Pretraining}\label{exp:multitask_train}
In the domain of audio processing, diverse audio datasets have been developed to address specific tasks, as shown in Table~\ref{tab:train_table}. Qwen-Audio aims to perform co-training using a wide range of audio datasets. The objective is to train a unified model capable of supporting all audio tasks, eliminating the need for laborious model switching when dealing with different tasks. More importantly, during co-training, tasks can benefit from each other since 1) similar tasks can benefit from knowledge sharing and collaborative learning, as they share a common focus on the fundamental information embedded within the audio signal; 2) tasks that rely on lower-level perceptual abilities can assist tasks that require higher-level understanding or reasoning capabilities.

However, different datasets exhibit significant variations in textual labels due to differences in task focus, language, granularity of annotation, and text structure (e.g., some data is structured while others are unstructured). To train a network for different tasks, simply mixing these diverse datasets cannot lead to mutual enhancement; instead, it introduces interference. Most existing multi-task training approaches have either grouped similar tasks (e.g., audio captioning, transcription) or assigned a dataset ID to each dataset~\citep{wang2023blsp, Macaw-LLM, NextGPT, LTU, shu2023llasm} to avoid interference. Although these approaches have achieved certain effectiveness, there is still considerable room for improvement. Whisper proposes a multitask training format by specifying tasks and condition information as a sequence of input special tokens to the language decoder, such as voice activity detection, language identification, and sentence-level timestamp tags. However, Whisper focuses on speech translation and recognition tasks only.

\begin{table}[t!]
\caption{Multi-task pre-training dataset.}
\centering
\vspace{-2mm}
\scalebox{1.0}{
\begin{tabular}{@{}lccc@{}}
\toprule
\textbf{Types} & \textbf{Task}                                               & \textbf{Description}                                                  & \textbf{Hours}         \\ \midrule
\multirow{18}{*}{\textbf{Speech}} & ASR      & Automatic speech recognition~(multiple languages)                                      & 30k           \\
 & S2TT                                & Speech-to-text translation                                                   & 3.7k          \\
 & OSR                                & Overlapped speech recognition                                                   & $<$1k          \\
 & Dialect ASR                                        & Automatic dialect speech recognition                                                                                 & 2k      \\
 & \multirow{2}{*}{SRWT} & English speech recognition with word-level timestamps                           & 10k          \\
&                       & Mandarin speech recognition with word-level timestamps                                           & 11k           \\
 & DID             &  Dialect identification     & 2k            \\
 & LID                          & Spoken language identification                             & 11.7k \\
 & SGC                        & Speaker gender recognition~(biologically)                                                                       & 4.8k \\
 & ER                                     & Emotion recognition                  & $<$1k  \\
 & SV                                & Speaker verification                  & 1.2k          \\
 & SD                              & Speaker diarization                         & $<$1k          \\
 & SER                    & Speech entity recognition                                     & $<$1k           \\
 & KS                      & Keyword spotting                                                                                                & $<$1k            \\
 & IC                      & Intent classification                                     & $<$1k            \\
 & SF                      & Slot filling                                            & $<$1k            \\
 & SAP                             & Speaker age prediction                                                         & 4.8k          \\
  & VSC              & Vocal sound classification                                                                  & $<$1k            \\\midrule

\multirow{5}{*}{\textbf{Sound}} & AAC                               & Automatic audio caption                           & 8.4k          \\
 & SEC                    & Sound event classification                   & 5.4k          \\
 & ASC                      & Acoustic scene classification    &  $<$1k          \\
 & SED                             & Sound event detection with timestamps                                        & $<$1k           \\
 & AQA                                    & Audio question answering                                                & $<$1k           \\\midrule

\multirow{9}{*}{\textbf{Music\&Song}} & SID     & Singer identification                                                                        & $<$1k           \\
 & SMER                        & Singer and music emotion recognition                                                                       & $<$1k \\
 & MC                & Music caption                     & 25k           \\
 & MIC                   & Music instruments classification                                              & $<$1k            \\
 & MNA                               & Music note analysis such as pitch, velocity &  $<$1k \\
 & MGR                            & Music genre recognition                                                       & 9.5k          \\
 & MR                          & Music recognition                                       & $<$1k           \\
 & MQA                                     & Music question answering                          & $<$1k               \\ \bottomrule

\end{tabular}%
}
\label{tab:train_table}
\end{table}

\paragraph{Multi-task Training Format Framework} Motivated by Whisper~\citep{Whisper}, to incorporate different kinds of audio, we propose a multitask training format framework as follows: 
\begin{itemize}
    \item Transcription Tag: The initiation of prediction is denoted using a transcription tag. The <|startoftranscripts|> is employed to indicate the tasks involve \textit{accurately} transcribing the spoken words and capturing the linguistic content of a speech recording, such as speech recognition and speech translation tasks. For other tasks, the <|startofanalysis|> tag is utilized.
    \item Audio Language Tag: Then, we incorporate a language tag that indicates the spoken language in the audio. This tag uses a unique token assigned to each language present in our training set, eight languages in totally. In the case where an audio segment does not contain any speech, such as natural sounds and music, the model is trained to predict a <|unknown|> token.
    \item Task Tag: The subsequent tokens specify the task. We categorize the collected audio tasks into five categories: <|transcribe|>, <|translate|>, <|caption|>, <|analysis|>, and <|question-answer|> tasks. For question-answer (QA) tasks, we append the corresponding questions after the tag.
    \item Text Language Tag: The tag token specifies the language of output text sequences.
    \item Timestamps Tag: The presence of a <|timestamps|> or <|notimestamps|> token determines whether the model needs to predict timestamps or not. Different from the sentence-level timestamps used in Whisper, the inclusion of the <|timestamps|> tag requires the model to perform fine-grained \textit{word-level} timestamp prediction, abbreviated as SRWT (Speech Recognition with Word-level Timestamps). The prediction of these timestamps is interleaved with the transcription words: the start time token is predicted before each transcription token, while the end time token is predicted after. According to our experiments, SRWT improves the ability of the model to align audio signals with timestamps. This improved alignment contributes to a comprehensive understanding of speech signals by the model, resulting in notable advancements across many tasks such as speech recognition and audio QA tasks.
    \item Output Instruction: Lastly, we provide output instruction to further specify the task and desired format for different subtasks, and then the text output begins.
\end{itemize}

The guiding principle behind our framework is to maximize the sharing of knowledge among similar tasks through shared tags, thereby improving their performance. Meanwhile, we ensure that different tasks and output formats can be distinguished to avoid the one-to-many mapping problem for the model. Please see Figure~\ref{img:format} for an overview of our multitask format of Qwen-Audio.

\subsection{Supervised Fine-tuning}\label{exp:sft}
The extensive pretraining of multitask models has equipped them with a broad understanding of audio. Building upon this, we employ instruction-based fine-tuning techniques to improve the ability of the model to align with human intent, resulting in an interactive chat model, termed Qwen-Audio-Chat. 
To accomplish this, we manually create demonstrations for each task. These demonstrations consist of raw text labels, questions, and answers. We then utilize GPT-3.5~\citep{chatgpt} to generate further questions and answers based on the provided raw text labels. Additionally, we also create a dataset of audio-dialogue data by employing manual annotation, model generation, and strategy concatenation. This dataset helps us incorporate reasoning, story generation, and multi-image comprehension abilities into our model. 

To handle multi-audio dialogue and multiple audio inputs effectively, we introduce the convention of labeling different audios with "Audio $id$:", where $id$ corresponds to the order of the audio input dialogue. In terms of dialogue format, we construct our instruction tuning dataset using the ChatML~\citep{chatml} format. In this format, each interaction's statement is marked with two special tokens ($<$im\_start$>$ and $<$im\_end$>$) to facilitate dialogue termination.

\begin{tcolorbox}[colback=black!5!white,colframe=black!75!black,title=The Data Format Example of Supervised Fine-Tuning.]
\textcolor{blue}{$<$im\_start$>$}user

Audio 1: $<$audio$>$emov-db/141-168-0155.wav$<$/audio$>$what does the speaker say?\textcolor{blue}{$<$im\_end$>$}

\textcolor{blue}{$<$im\_start$>$}assistant

\textcolor{blue}{The speaker says in English, "Won't you draw up, gentlemen.".$<$im\_end$>$}

\textcolor{blue}{$<$im\_start$>$}user

What's the mood of the person?\textcolor{blue}{$<$im\_end$>$}

\textcolor{blue}{$<$im\_start$>$}assistant

\textcolor{blue}{Based on the voice, the mood of the person is disgusted.$<$im\_end$>$}
\end{tcolorbox}

In order to facilitate versatile input from both audio and pure text modalities within multi-turn dialogues, we use a combination of audio-centric instruction data mentioned above and pure text instruction data during this training process. This approach allows the model to handle diverse forms of input seamlessly. The total amount of instruction tuning data is 20k.

% During training, we ensure the consistency between prediction and training distributions by only supervising answers and special tokens (blue in the example), and not supervising role names or question prompts.  

\begin{table}[t!]
\centering
\caption{Summary of the evaluation benchmarks for Qwen-Audio.}
\vspace{-2mm}
\resizebox{\textwidth}{!}{%
\begin{tabular}{lcccc}
\toprule
\textbf{Task}                 & \textbf{Description}                                & \textbf{Dataset}     & \textbf{Split}      & \textbf{Metric}                                  \\\midrule
\multirow{3}{*}{ASR} & \multirow{3}{*}{Automatic Speech Recognition} & Aishell1~\citep{aishell1}    & dev | test & \multirow{3}{*}{WER}                    \\
                     &                                               & Aishell2~\citep{aishell2}    & test        &                                         \\
                     &                                               & Librispeech~\citep{Librispeech} & dev | test &                                         \\\midrule
S2TT                  & Speech-to-text translation                           & CoVoST2~\citep{CoVoST2}     & test        & BLEU\footnote~\citep{papineni2002bleu}                                   \\\midrule
SRWT                & Speech Recognition with Word-level Timestamp                     &      Industrial Data~\citep{FunASR}    & test        & AAS~\citep{shi2023achieving}                                     \\\midrule
\multirow{2}{*}{AAC} & \multirow{2}{*}{Automatic Audio Caption}      & \multirow{2}{*}{Clotho~\citep{Clotho}}      &     \multirow{2}{*}{test}        & CIDEr | SPICE | SPIDEr \\
 &       &      &           & \citep{vedantam2015cider, anderson2016spice} \\

\midrule
\multirow{2}{*}{ASC} & \multirow{2}{*}{Acoustic Scene Classfication} & CochlScene~\citep{cochlscene}  &     test        & \multirow{2}{*}{ACC}                    \\
                     &                                               & TUT2017~\citep{TUT2017}     &    eval         &                                         \\\midrule
SER                  & Speech Emotion Recognition                    & Meld~\citep{Meld}        &      test       & ACC                                     \\\midrule
AQA                  & Audio Question \& Answer                      & ClothoAQA~\citep{ClothoAQA}   &     test        & ACC                                     \\\midrule
VSC                  & Vocal Sound Classification                     & VocalSound~\citep{VocalSound}  &    test         & ACC                                     \\\midrule
MNA                  & Music Note Analysis                           & NSynth~\citep{NSynth}      &       test      & ACC / MAP                      \\\bottomrule         
\end{tabular}
}
\label{tab:evaluation}
\end{table}
\footnotetext{https://github.com/mjpost/sacrebleu}

\section{Experiments}
\label{sec:experiments}
\subsection{Setup}
For multi-task pre-training, we freeze the weights of LLM and only optimize the audio encoder. This trained model is referred to as Qwen-Audio. In the subsequent supervised fine-tuning stage, we fix the weights of the audio encoder and only optimize the LLM. The resulting model is denoted as Qwen-Audio-Chat. The detailed training configurations of both stages are listed in Table~\ref{tab:hyperparam}.

\subsection{Evaluation}

In order to assess the universal understanding capabilities of Qwen-Audio, as shown in Table~\ref{tab:evaluation}, we perform a comprehensive evaluation that encompasses various tasks, namely Automatic Speech Recognition (ASR), Speech-to-Text Translation (S2TT), Automatic Audio Captioning (AAC), Acoustic Scene Classification (ASC), Speech Emotion Recognition (SER), Audio Question and Answering (AQA), Vocal Sound Classification (VSC), and Music Note Analysis (MNA). This evaluation is conducted across 12 datasets. The evaluation datasets are rigorously excluded from the training data to avoid data leakage. The detailed training configurations of both stages are listed in Table~\ref{tab:hyperparam}.

\begin{table}[t!]
\centering
\caption{The results of Automatic Speech Recognition~(ASR), Speech-to-Text Translation~(S2TT), Automatic Audio Captioning~(AAC), Speech Recognition with Word-level Timestamps~(SRWT), Acoustic Scene Classification~(ASC), Speech Emotion Recognition~(SER), Audio Question and Answering~(AQA), Vocal Sound Classification~(VSC), and Music Note Analysis~(MNA) tasks. For SRWT task, the results of Forced-aligner~\citep{forcealigner} are to predict the timestamps given the ground-truth transcripts, while Paraformer-large-TP~\citep{FunASR} and Qwen-audio tackle a more challenging scenario by directly generating sequences containing both transcriptions and timestamps.}
\vspace{-2mm}
\resizebox{\textwidth}{!}{%
\begin{tabular}{lcccc}
\toprule
\multirow{2}{*}{\textbf{Task}}                           & \multirow{2}{*}{\textbf{Dataset} }       & \multirow{2}{*}{\textbf{Model}} & \multicolumn{2}{c}{\textbf{Performance} }                                \\ \cmidrule(l){4-5} 
                                                  &                                 &                        & \textbf{Metrics}                                 & \textbf{Results}               \\ \midrule
\multirow{12}{*}{ASR}    & \multirow{5}{*}{\begin{tabular}[c]{@{}c@{}}\textbf{Librispeech}\\      \textit{dev-clean} | \textit{dev-other} | \\ \textit{test-clean} | \textit{test-other} \end{tabular}}     & SpeechT5~\citep{ao2021speecht5}    & \multirow{5}{*}{WER~$\downarrow$}                  & 2.1 | 5.5 | 2.4 | 5.8                \\
                                                  &                                  & SpeechNet~\citep{chen2021speechnet}               &                                         & - | - | 30.7 | -            \\
                                                  &                              & SLM-FT~\citep{SLM}               &                                         & - | - | 2.6 | 5.0            \\
                                                  
                                                  &                           & SALMONN~\citep{anonymous2023salmonn}               &                                         & - | - | 2.1 | 4.9            \\
                                                  
                                                  &                                   & Qwen-Audio               &                                         & \textbf{1.8} | \textbf{4.0} | \textbf{2.0} | \textbf{4.2}            \\
\cmidrule(l){2-5}
                                                  & \multirow{4}{*}{\begin{tabular}[c]{@{}c@{}}\textbf{Aishell1} \\ \textit{dev} | \textit{test} \end{tabular}}        & MMSpeech-base~\citep{mmspeech}              &       \multirow{4}{*}{WER~$\downarrow$}            & 2.0 | 2.1                \\
                                                  &                           & MMSpeech-large~\citep{mmspeech}                  &                                         & 1.6 | 1.9 
                                                  \\
                                                  &                           & Paraformer-large~\citep{FunASR}          &                                         & - | 2.0                 \\
                                                  
                                                  &                           & Qwen-Audio           &                                         & \textbf{1.2} | \textbf{1.3}                 \\
\cmidrule(l){2-5} 
                                                  & \multirow{3}{*}{\begin{tabular}[c]{@{}c@{}}\textbf{Aishell2} \\ \textit{Mic} | \textit{iOS} | \textit{Android} \end{tabular}}        & MMSpeech-base~\citep{mmspeech}              &     \multirow{3}{*}{WER~$\downarrow$}              & 4.5 | 3.9 | 4.0                \\
                                                  &                           & Paraformer-large~\citep{FunASR}          &                                         & - | \textbf{2.9} | -                 \\
                                                  
                                                  &                           & Qwen-Audio           &                                         & \textbf{3.3} | 3.1 | \textbf{3.3}                 \\
                                                  \midrule
\multirow{6}{*}{S2TT}    & \multirow{4}{*}{\begin{tabular}[c]{@{}c@{}}\textbf{CoVoST2} \\ \textit{en-de} | \textit{de-en} | \\ \textit{en-zh} | \textit{zh-en} \end{tabular}}     & SALMONN~\citep{anonymous2023salmonn}    & \multirow{4}{*}{BLEU~$\uparrow$}                    & 18.6 | - | 33.1 | -                \\
                                                  &                              & SpeechLLaMA~\citep{speechllama}                &                                         & - | 27.1 | - | 12.3            \\
                                                  
                                                  &                         & BLSP~\citep{wang2023blsp}            &                                         & 14.1 | - | - | -            \\
                                                  &                               & Qwen-Audio                &                                         & \textbf{25.1} | \textbf{33.9} | \textbf{41.5} | \textbf{15.7}            \\
\cmidrule(l){2-5}
                                                  & \textbf{CoVoST2}        & SpeechLLaMA~\citep{speechllama}              &    \multirow{2}{*}{BLEU~$\uparrow$}              & 27.9 | 25.2 | 25.9                \\
                                                  &      \textit{es-en} | \textit{fr-en} | \textit{it-en}                           & Qwen-Audio                   &                                         & \textbf{39.7} | \textbf{38.5} | \textbf{36.0}                 \\ \midrule
                                                  
\multirow{2}{*}{AAC}    & \multirow{2}{*}{\textbf{Clotho}}         & Pengi~\citep{Pengi}                 & \multirow{2}{*}{\begin{tabular}[c]{@{}c@{}}CIDEr | \\ SPICE | SPIDEr~$\uparrow$ \end{tabular}}
& 0.416 | 0.126 | 0.271 \\
                                                  &                                 & Qwen-Audio                   &                                         & \textbf{0.441} | \textbf{0.136} | \textbf{0.288}  \\ \midrule
                                    
\multirow{3}{*}{SRWT}   & \multirow{3}{*}{\textbf{Industrial Data} }        & Force-aligner~\citep{forcealigner}               & \multirow{3}{*}{AAS~(ms)~$\downarrow$} & 60.3 \\
     &           & Paraformer-large-TP~\citep{FunASR}                   &                                         & 65.3  \\
      &                    & Qwen-Audio                   &                                         & \textbf{51.5}  \\\midrule
\multirow{4}{*}{ASC}    & \multirow{2}{*}{\textbf{CochlScene}}     & CochlScene~\citep{cochlscene}    & \multirow{2}{*}{ACC~$\uparrow$}                    & 0.669                 \\
                                                  &                                 & Qwen-Audio                   &                                         & \textbf{0.795}                 \\ \cmidrule(l){2-5} 
                                                  & \multirow{2}{*}{\textbf{TUT2017} }       & Pengi~\citep{Pengi}                  & \multirow{2}{*}{ACC~$\uparrow$}                    & 0.353                 \\
                                                  &                                 & Qwen-Audio                   &                                         & \textbf{0.649}                 \\ \midrule
\multirow{2}{*}{SER} & \multirow{2}{*}{\textbf{Meld}}           & WavLM-large~\citep{wavlm}            & \multirow{2}{*}{ACC~$\uparrow$}                    & 0.542                 \\
                                                  &                                 & Qwen-Audio                    &                                         & \textbf{0.557}                 \\ \midrule
\multirow{3}{*}{AQA}   & \multirow{3}{*}{\textbf{ClothoAQA}}      & ClothoAQA~\citep{ClothoAQA}     & \multirow{3}{*}{ACC | ACC (binary)~$\uparrow$}     & 0.542 | 0.627         \\
                                                  &                                 & Pengi~\citep{Pengi}                  &                                         & - | 0.645             \\
                                                  &                                 & Qwen-Audio                    &                                         & \textbf{0.579} | \textbf{0.749}         \\ \midrule
\multirow{3}{*}{VSC}       & \multirow{3}{*}{\textbf{VocalSound}}     & CLAP~\citep{CLAP}                   & \multirow{3}{*}{ACC~$\uparrow$}                    & 0.4945                \\
                                                  &                                 & Pengi~\citep{Pengi}                  &                                         & 0.6035                \\
                                                  &                                 & Qwen-Audio                    &                                         & \textbf{0.9289}                \\ \midrule
\multirow{4}{*}{MNA}             
                                                  & \multirow{2}{*}{\textbf{NS. Qualities}}  & Pengi~\citep{Pengi}                  & \multirow{2}{*}{MAP~$\uparrow$}                    & 0.3860                \\
                                                  &                                 & Qwen-Audio                    &                                         & \textbf{0.4742}                \\ \cmidrule(l){2-5} 
                                                  & \multirow{2}{*}{\textbf{NS. Instrument}} & Pengi~\citep{Pengi}                  & \multirow{2}{*}{ACC~$\uparrow$}                    & 0.5007                \\
                                                  &                                 & Qwen-Audio                   &                                         & \textbf{0.7882}                \\ \bottomrule
\end{tabular}%
}
\label{tab:audio_analysis_table}
\end{table}

\subsection{Main Results}
In this section, we present a comprehensive evaluation of the Qwen-Audio model, assessing its performance across various tasks without any task-specific fine-tuning. We begin by examining its English Automatic Speech Recognition (ASR) results, as depicted in Table~\ref{tab:audio_analysis_table}, where Qwen-Audio exhibits superior performance compared to previous multi-task learning models. Specifically, it achieves a 2.0\% and 4.2\% WER on the librispeech test-clean and test-other datasets, respectively. Similarly, the Chinese Mandarin ASR results demonstrate Qwen-Audio's competitive performance against previous approaches. To the best of our knowledge, Qwen-Audio achieves state-of-the-art results on the Aishell1 dev and test sets. Furthermore, we evaluate Qwen-Audio's speech translation performance on the CoVoST2 dataset. The results reveal that Qwen-Audio outperforms the baselines by a substantial margin across all seven translation directions.

Lastly, we analyze the performance of Qwen-Audio on various audio analysis tasks, including AAC, SWRT ASC, SER, AQA, VSC, and MNA, as summarized in Table~\ref{tab:audio_analysis_table}. Across these tasks, Qwen-Audio consistently outperforms the baselines by a significant margin. Notably, it achieves state-of-the-art results on CochlScene, ClothoAQA, and VocalSound, thereby demonstrating the model's robust audio understanding capabilities.

\subsection{Results of Interactive Chat}
We showcase the conversational capabilities of Qwen-Audio-Chat through illustrative cases depicted in Figure~\ref{img:demo}. Furthermore, we intend to provide public access to the trained models for online chat interactions.

\subsection{The Analysis of Word-level Timestamps Prediction}
We propose the task of speech recognition with word-level timestamps (SRWT) by training Qwen-Audio to not only recognize speech transcripts but also predict the timestamps for each word. The purpose of SRWT is twofold: firstly, to improve the model's ability to align audio signals with fine-grained timestamps; secondly, to support grounding of speech and audio, and grounding-based QA tasks in Qwen-Audio-Chat, such as finding the starting and ending time of an audio segment mentioning a person's name or identifying whether a sound occurs in the given audio~\footnote{Audio event detection can be considered as a subtask of event timestamp prediction as the absence of an event's timestamp implies its non-occurrence in the audio.}. 

In this section, we exclude the training of SRWT tasks from multi-task pretraining while maintaining the other tasks unchanged. Notably, the removal of SRWT does not impact the coverage of audio datasets for training since SRWT tasks share the same audio dataset as automatic speech recognition (ASR) tasks. The results are shown in Table~\ref{tab:ab_asr} and Table~\ref{tab:ab_aqa}: models trained with SRWT achieve superior performance in automatic speech recognition and audio question-answering tasks, including natural sounds QA and Music QA. These results highlight the efficacy of incorporating fine-grained word-level timestamps to enhance the general audio signal grounding ability and subsequently improve the performance of sound and music signal QA tasks.

\begin{table}[t!]
\centering
\caption{Results of ASR tasks with or without training word-level timestamps tasks.}
\vspace{-2mm}
\begin{tabular}{lcccccc}
\toprule
\multirow{2}{*}{\textbf{Method}} & \multicolumn{4}{c}{\textbf{LibriSpeech} }                              & \multicolumn{2}{c}{\textbf{AISHELL1}}  \\ \cmidrule(l){2-7} 
                  & dev-clean     & dev-other     & test-clean    & test-other    & dev           & test          \\ \midrule
w/o SRWT & 1.93          & 4.18          & 2.22          & 4.21          & 1.54          & 1.71          \\
Qwen-Audio               & \textbf{1.79} & \textbf{4.00} & \textbf{2.04} & \textbf{4.19} & \textbf{1.22} & \textbf{1.29} \\ \bottomrule
\end{tabular}%
\label{tab:ab_asr}
\end{table}

\begin{table}[t!]
\centering
\caption{Results of AQA tasks with or without training word-level timestamps tasks.}
\vspace{-2mm}
\begin{tabular}{lccc}
\toprule
\multirow{2}{*}{\textbf{Method}} & \multicolumn{2}{c}{\textbf{ClothoAQA}}     & \textbf{MusicAVQA}       \\ \cmidrule(l){2-4} 
                  & test            & test-binary     & audio question  \\ \midrule
w/o SRWT & 0.5648          & 0.7418          & 0.7027          \\
Qwen-Audio      & \textbf{0.5795} & \textbf{0.7491} & \textbf{0.7211} \\ \bottomrule
\end{tabular}%
\label{tab:ab_aqa}
\end{table}

\section{Conclusion}
\label{sec:conclusion}
In this paper, we present the Qwen-Audio series, a set of large-scale audio-language models with universal audio understanding abilities. To incorporate different kinds of audios for co-training, we propose a unified multi-task learning framework that facilitates the sharing of knowledge among similar tasks and avoids one-to-many mapping problem caused by different text formats. Without any task-specific fine-tuning, the resulting Qwen-Audio models outperform previous works across diverse benchmarks, demonstrating its universal audio understanding abilities. Through supervised instruction finetuning, Qwen-Audio-Chat showcases robust capabilities in aligning with human intent, supporting multilingual and multi-turn dialogues from both audio and text inputs.

\section{Acknowledgements}
We express our gratitude to Jinze Bai, Shuai Bai, Peng Wang, Sinan Tan, Shijie Wang for their insightful discussion. We would like to thank Juan Zhu, Junyang Lin, Siqi Zheng, Jiaming Wang and Zhihao Du for their support of this project.

\bibliographystyle{plainnat} 
\bibliography{references}

\begin{thebibliography}{71}
\providecommand{\natexlab}[1]{#1}
\providecommand{\url}[1]{\texttt{#1}}
\expandafter\ifx\csname urlstyle\endcsname\relax
  \providecommand{\doi}[1]{doi: #1}\else
  \providecommand{\doi}{doi: \begingroup \urlstyle{rm}\Url}\fi

\bibitem[Alayrac et~al.(2022)Alayrac, Donahue, Luc, Miech, Barr, Hasson, Lenc, Mensch, Millican, Reynolds, et~al.]{alayrac2022flamingo}
Jean-Baptiste Alayrac, Jeff Donahue, Pauline Luc, Antoine Miech, Iain Barr, Yana Hasson, Karel Lenc, Arthur Mensch, Katherine Millican, Malcolm Reynolds, et~al.
\newblock Flamingo: a visual language model for few-shot learning.
\newblock \emph{NeurIPS}, 2022.

\bibitem[Anderson et~al.(2016)Anderson, Fernando, Johnson, and Gould]{anderson2016spice}
Peter Anderson, Basura Fernando, Mark Johnson, and Stephen Gould.
\newblock Spice: Semantic propositional image caption evaluation.
\newblock In \emph{Computer Vision--ECCV 2016: 14th European Conference, Amsterdam, The Netherlands, October 11-14, 2016, Proceedings, Part V 14}. Springer, 2016.

\bibitem[Anil et~al.(2023)Anil, Dai, Firat, Johnson, Lepikhin, Passos, Shakeri, Taropa, Bailey, Chen, et~al.]{palm2}
Rohan Anil, Andrew~M Dai, Orhan Firat, Melvin Johnson, Dmitry Lepikhin, Alexandre Passos, Siamak Shakeri, Emanuel Taropa, Paige Bailey, Zhifeng Chen, et~al.
\newblock {PaLM} 2 technical report.
\newblock \emph{arXiv:2305.10403}, 2023.

\bibitem[Anonymous(2023)]{anonymous2023salmonn}
Anonymous.
\newblock {SALMONN}: Towards generic hearing abilities for large language models.
\newblock In \emph{Submitted to The Twelfth International Conference on Learning Representations}, 2023.
\newblock under review.

\bibitem[Ao et~al.(2021)Ao, Wang, Zhou, Wang, Ren, Wu, Liu, Ko, Li, Zhang, et~al.]{ao2021speecht5}
Junyi Ao, Rui Wang, Long Zhou, Chengyi Wang, Shuo Ren, Yu~Wu, Shujie Liu, Tom Ko, Qing Li, Yu~Zhang, et~al.
\newblock Speecht5: Unified-modal encoder-decoder pre-training for spoken language processing.
\newblock \emph{arXiv:2110.07205}, 2021.

\bibitem[Bai et~al.(2023{\natexlab{a}})Bai, Bai, Chu, Cui, Dang, Deng, Fan, Ge, Han, Huang, et~al.]{bai2023qwen}
Jinze Bai, Shuai Bai, Yunfei Chu, Zeyu Cui, Kai Dang, Xiaodong Deng, Yang Fan, Wenbin Ge, Yu~Han, Fei Huang, et~al.
\newblock Qwen technical report.
\newblock \emph{arXiv preprint arXiv:2309.16609}, 2023{\natexlab{a}}.

\bibitem[Bai et~al.(2023{\natexlab{b}})Bai, Bai, Yang, Wang, Tan, Wang, Lin, Zhou, and Zhou]{qwenvl}
Jinze Bai, Shuai Bai, Shusheng Yang, Shijie Wang, Sinan Tan, Peng Wang, Junyang Lin, Chang Zhou, and Jingren Zhou.
\newblock {Qwen-VL}: {A} frontier large vision-language model with versatile abilities.
\newblock \emph{CoRR}, abs/2308.12966, 2023{\natexlab{b}}.

\bibitem[Brown et~al.(2020)Brown, Mann, Ryder, Subbiah, Kaplan, Dhariwal, Neelakantan, Shyam, Sastry, Askell, et~al.]{gpt3}
Tom Brown, Benjamin Mann, Nick Ryder, Melanie Subbiah, Jared~D Kaplan, Prafulla Dhariwal, Arvind Neelakantan, Pranav Shyam, Girish Sastry, Amanda Askell, et~al.
\newblock Language models are few-shot learners.
\newblock \emph{NeurIPS}, 2020.

\bibitem[Bu et~al.(2017)Bu, Du, Na, Wu, and Zheng]{aishell1}
Hui Bu, Jiayu Du, Xingyu Na, Bengu Wu, and Hao Zheng.
\newblock {AISHELL-1:} an open-source mandarin speech corpus and a speech recognition baseline.
\newblock In \emph{20th Conference of the Oriental Chapter of the International Coordinating Committee on Speech Databases and Speech {I/O} Systems and Assessment, {O-COCOSDA} 2017, Seoul, South Korea, November 1-3, 2017}. {IEEE}, 2017.

\bibitem[Chen et~al.(2023)Chen, Zhang, Zeng, Zhang, Zhu, and Zhao]{shikra}
Keqin Chen, Zhao Zhang, Weili Zeng, Richong Zhang, Feng Zhu, and Rui Zhao.
\newblock Shikra: Unleashing multimodal llm's referential dialogue magic.
\newblock \emph{arXiv:2306.15195}, 2023.

\bibitem[Chen et~al.(2022)Chen, Wang, Chen, Wu, Liu, Chen, Li, Kanda, Yoshioka, Xiao, Wu, Zhou, Ren, Qian, Qian, Wu, Zeng, Yu, and Wei]{wavlm}
Sanyuan Chen, Chengyi Wang, Zhengyang Chen, Yu~Wu, Shujie Liu, Zhuo Chen, Jinyu Li, Naoyuki Kanda, Takuya Yoshioka, Xiong Xiao, Jian Wu, Long Zhou, Shuo Ren, Yanmin Qian, Yao Qian, Jian Wu, Michael Zeng, Xiangzhan Yu, and Furu Wei.
\newblock Wavlm: Large-scale self-supervised pre-training for full stack speech processing.
\newblock \emph{{IEEE} J. Sel. Top. Signal Process.}, 2022.

\bibitem[Chen et~al.(2021)Chen, Chi, Yang, Chang, Lin, Huang, Liu, Liu, Lee, and Lee]{chen2021speechnet}
Yi-Chen Chen, Po-Han Chi, Shu-wen Yang, Kai-Wei Chang, Jheng-hao Lin, Sung-Feng Huang, Da-Rong Liu, Chi-Liang Liu, Cheng-Kuang Lee, and Hung-yi Lee.
\newblock Speechnet: A universal modularized model for speech processing tasks.
\newblock \emph{arXiv:2105.03070}, 2021.

\bibitem[Chowdhery et~al.(2022)Chowdhery, Narang, Devlin, Bosma, Mishra, Roberts, Barham, Chung, Sutton, Gehrmann, et~al.]{palm}
Aakanksha Chowdhery, Sharan Narang, Jacob Devlin, Maarten Bosma, Gaurav Mishra, Adam Roberts, Paul Barham, Hyung~Won Chung, Charles Sutton, Sebastian Gehrmann, et~al.
\newblock {PaLM}: Scaling language modeling with pathways.
\newblock \emph{arXiv:2204.02311}, 2022.

\bibitem[D{\'e}fossez et~al.(2022)D{\'e}fossez, Copet, Synnaeve, and Adi]{Encodec}
Alexandre D{\'e}fossez, Jade Copet, Gabriel Synnaeve, and Yossi Adi.
\newblock High fidelity neural audio compression.
\newblock \emph{arXiv:2210.13438}, 2022.

\bibitem[Deshmukh et~al.(2023)Deshmukh, Elizalde, Singh, and Wang]{Pengi}
Soham Deshmukh, Benjamin Elizalde, Rita Singh, and Huaming Wang.
\newblock Pengi: An audio language model for audio tasks.
\newblock \emph{CoRR}, 2023.

\bibitem[Drossos et~al.(2020)Drossos, Lipping, and Virtanen]{Clotho}
Konstantinos Drossos, Samuel Lipping, and Tuomas Virtanen.
\newblock Clotho: an audio captioning dataset.
\newblock In \emph{2020 {IEEE} International Conference on Acoustics, Speech and Signal Processing, {ICASSP} 2020, Barcelona, Spain, May 4-8, 2020}. {IEEE}, 2020.

\bibitem[Du et~al.(2018)Du, Na, Liu, and Bu]{aishell2}
Jiayu Du, Xingyu Na, Xuechen Liu, and Hui Bu.
\newblock {AISHELL-2:} transforming mandarin {ASR} research into industrial scale.
\newblock abs/1808.10583, 2018.

\bibitem[Elizalde et~al.(2022)Elizalde, Deshmukh, Ismail, and Wang]{CLAP}
Benjamin Elizalde, Soham Deshmukh, Mahmoud~Al Ismail, and Huaming Wang.
\newblock {CLAP:} learning audio concepts from natural language supervision.
\newblock abs/2206.04769, 2022.

\bibitem[Engel et~al.(2017)Engel, Resnick, Roberts, Dieleman, Norouzi, Eck, and Simonyan]{NSynth}
Jesse~H. Engel, Cinjon Resnick, Adam Roberts, Sander Dieleman, Mohammad Norouzi, Douglas Eck, and Karen Simonyan.
\newblock Neural audio synthesis of musical notes with wavenet autoencoders.
\newblock In \emph{Proceedings of the 34th International Conference on Machine Learning, {ICML} 2017, Sydney, NSW, Australia, 6-11 August 2017}, Proceedings of Machine Learning Research. {PMLR}, 2017.

\bibitem[Gao et~al.(2023)Gao, Li, Wang, Luo, Shi, Chen, Li, Zuo, Du, Xiao, and Zhang]{FunASR}
Zhifu Gao, Zerui Li, Jiaming Wang, Haoneng Luo, Xian Shi, Mengzhe Chen, Yabin Li, Lingyun Zuo, Zhihao Du, Zhangyu Xiao, and Shiliang Zhang.
\newblock Funasr: {A} fundamental end-to-end speech recognition toolkit.
\newblock \emph{CoRR}, abs/2305.11013, 2023.

\bibitem[Gong et~al.(2022)Gong, Yu, and Glass]{VocalSound}
Yuan Gong, Jin Yu, and James~R. Glass.
\newblock Vocalsound: {A} dataset for improving human vocal sounds recognition.
\newblock In \emph{{IEEE} International Conference on Acoustics, Speech and Signal Processing, {ICASSP} 2022, Virtual and Singapore, 23-27 May 2022}, pages 151--155. {IEEE}, 2022.
\newblock \doi{10.1109/ICASSP43922.2022.9746828}.
\newblock URL \url{https://doi.org/10.1109/ICASSP43922.2022.9746828}.

\bibitem[Gong et~al.(2023{\natexlab{a}})Gong, Khurana, Karlinsky, and Glass]{whisperAT}
Yuan Gong, Sameer Khurana, Leonid Karlinsky, and James~R. Glass.
\newblock Whisper-at: Noise-robust automatic speech recognizers are also strong general audio event taggers.
\newblock \emph{CoRR}, abs/2307.03183, 2023{\natexlab{a}}.

\bibitem[Gong et~al.(2023{\natexlab{b}})Gong, Luo, Liu, Karlinsky, and Glass]{LTU}
Yuan Gong, Hongyin Luo, Alexander~H. Liu, Leonid Karlinsky, and James~R. Glass.
\newblock Listen, think, and understand.
\newblock \emph{CoRR}, abs/2305.10790, 2023{\natexlab{b}}.

\bibitem[Hsu et~al.(2021)Hsu, Bolte, Tsai, Lakhotia, Salakhutdinov, and Mohamed]{HuBERT}
Wei{-}Ning Hsu, Benjamin Bolte, Yao{-}Hung~Hubert Tsai, Kushal Lakhotia, Ruslan Salakhutdinov, and Abdelrahman Mohamed.
\newblock Hubert: Self-supervised speech representation learning by masked prediction of hidden units.
\newblock \emph{{IEEE} {ACM} Trans. Audio Speech Lang. Process.}, 2021.

\bibitem[Hu et~al.(2021)Hu, Shen, Wallis, Allen-Zhu, Li, Wang, Wang, and Chen]{hu2021lora}
Edward~J Hu, Yelong Shen, Phillip Wallis, Zeyuan Allen-Zhu, Yuanzhi Li, Shean Wang, Lu~Wang, and Weizhu Chen.
\newblock Lora: Low-rank adaptation of large language models.
\newblock \emph{arXiv:2106.09685}, 2021.

\bibitem[Huang et~al.(2023)Huang, Li, Yang, Shi, Chang, Ye, Wu, Hong, Huang, Liu, Ren, Zhao, and Watanabe]{AudioGPT}
Rongjie Huang, Mingze Li, Dongchao Yang, Jiatong Shi, Xuankai Chang, Zhenhui Ye, Yuning Wu, Zhiqing Hong, Jiawei Huang, Jinglin Liu, Yi~Ren, Zhou Zhao, and Shinji Watanabe.
\newblock Audiogpt: Understanding and generating speech, music, sound, and talking head.
\newblock \emph{CoRR}, abs/2304.12995, 2023.

\bibitem[Jeong and Park(2022)]{cochlscene}
Il{-}Young Jeong and Jeongsoo Park.
\newblock Cochlscene: Acquisition of acoustic scene data using crowdsourcing.
\newblock abs/2211.02289, 2022.

\bibitem[Le et~al.(2023)Le, Vyas, Shi, Karrer, Sari, Moritz, Williamson, Manohar, Adi, Mahadeokar, and Hsu]{VoiceBox}
Matthew Le, Apoorv Vyas, Bowen Shi, Brian Karrer, Leda Sari, Rashel Moritz, Mary Williamson, Vimal Manohar, Yossi Adi, Jay Mahadeokar, and Wei{-}Ning Hsu.
\newblock Voicebox: Text-guided multilingual universal speech generation at scale.
\newblock \emph{CoRR}, 2023.

\bibitem[Li et~al.(2022)Li, Li, Xiong, and Hoi]{blip}
Junnan Li, Dongxu Li, Caiming Xiong, and Steven C.~H. Hoi.
\newblock Blip: Bootstrapping language-image pre-training for unified vision-language understanding and generation.
\newblock In \emph{ICML}, 2022.

\bibitem[Li et~al.(2023)Li, Li, Savarese, and Hoi]{blip2}
Junnan Li, Dongxu Li, Silvio Savarese, and Steven C.~H. Hoi.
\newblock {BLIP-2:} bootstrapping language-image pre-training with frozen image encoders and large language models.
\newblock In \emph{International Conference on Machine Learning, {ICML} 2023, 23-29 July 2023, Honolulu, Hawaii, {USA}}, Proceedings of Machine Learning Research. {PMLR}, 2023.

\bibitem[Lipping et~al.(2022)Lipping, Sudarsanam, Drossos, and Virtanen]{ClothoAQA}
Samuel Lipping, Parthasaarathy Sudarsanam, Konstantinos Drossos, and Tuomas Virtanen.
\newblock Clotho-aqa: {A} crowdsourced dataset for audio question answering.
\newblock In \emph{30th European Signal Processing Conference, {EUSIPCO} 2022, Belgrade, Serbia, August 29 - Sept. 2, 2022}. {IEEE}, 2022.

\bibitem[Lyu et~al.(2023)Lyu, Wu, Wang, Huang, Liu, Du, Shi, and Tu]{Macaw-LLM}
Chenyang Lyu, Minghao Wu, Longyue Wang, Xinting Huang, Bingshuai Liu, Zefeng Du, Shuming Shi, and Zhaopeng Tu.
\newblock Macaw-llm: Multi-modal language modeling with image, audio, video, and text integration.
\newblock \emph{CoRR}, abs/2306.09093, 2023.

\bibitem[Maiti et~al.(2023)Maiti, Peng, Choi, Jung, Chang, and Watanabe]{maiti2023voxtlm}
Soumi Maiti, Yifan Peng, Shukjae Choi, Jee-weon Jung, Xuankai Chang, and Shinji Watanabe.
\newblock Voxtlm: unified decoder-only models for consolidating speech recognition/synthesis and speech/text continuation tasks.
\newblock \emph{arXiv:2309.07937}, 2023.

\bibitem[McAuliffe et~al.(2017)McAuliffe, Socolof, Mihuc, Wagner, and Sonderegger]{forcealigner}
Michael McAuliffe, Michaela Socolof, Sarah Mihuc, Michael Wagner, and Morgan Sonderegger.
\newblock Montreal forced aligner: Trainable text-speech alignment using kaldi.
\newblock In \emph{Interspeech 2017, 18th Annual Conference of the International Speech Communication Association, Stockholm, Sweden, August 20-24, 2017}, 2017.

\bibitem[Mesaros et~al.(2017)Mesaros, Heittola, Diment, Elizalde, Shah, Vincent, Raj, and Virtanen]{TUT2017}
Annamaria Mesaros, Toni Heittola, Aleksandr Diment, Benjamin Elizalde, Ankit Shah, Emmanuel Vincent, Bhiksha Raj, and Tuomas Virtanen.
\newblock {DCASE2017} challenge setup: Tasks, datasets and baseline system.
\newblock In \emph{Proceedings of the Workshop on Detection and Classification of Acoustic Scenes and Events, {DCASE} 2017, Munich, Germany, November 16-17, 2017}, 2017.

\bibitem[Nachmani et~al.(2023)Nachmani, Levkovitch, Salazar, Asawaroengchai, Mariooryad, Skerry{-}Ryan, and Ramanovich]{LMWithVoice}
Eliya Nachmani, Alon Levkovitch, Julian Salazar, Chulayuth Asawaroengchai, Soroosh Mariooryad, R.~J. Skerry{-}Ryan, and Michelle~Tadmor Ramanovich.
\newblock Lms with a voice: Spoken language modeling beyond speech tokens.
\newblock \emph{CoRR}, 2023.

\bibitem[Openai()]{chatml}
Openai.
\newblock Chatml documents.
\newblock URL \url{https://github.com/openai/openai-python/blob/main/chatml.md}.

\bibitem[OpenAI(2022)]{chatgpt}
OpenAI.
\newblock Introducing {ChatGPT}, 2022.
\newblock URL \url{https://openai.com/blog/chatgpt}.

\bibitem[OpenAI(2023)]{gpt4}
OpenAI.
\newblock Gpt-4 technical report, 2023.

\bibitem[Ouyang et~al.(2022)Ouyang, Wu, Jiang, Almeida, Wainwright, Mishkin, Zhang, Agarwal, Slama, Ray, et~al.]{ouyang2022intructgpt}
Long Ouyang, Jeffrey Wu, Xu~Jiang, Diogo Almeida, Carroll Wainwright, Pamela Mishkin, Chong Zhang, Sandhini Agarwal, Katarina Slama, Alex Ray, et~al.
\newblock Training language models to follow instructions with human feedback.
\newblock \emph{NeurIPS}, 2022.

\bibitem[Panayotov et~al.(2015)Panayotov, Chen, Povey, and Khudanpur]{Librispeech}
Vassil Panayotov, Guoguo Chen, Daniel Povey, and Sanjeev Khudanpur.
\newblock Librispeech: An {ASR} corpus based on public domain audio books.
\newblock In \emph{2015 {IEEE} International Conference on Acoustics, Speech and Signal Processing, {ICASSP} 2015, South Brisbane, Queensland, Australia, April 19-24, 2015}. {IEEE}, 2015.

\bibitem[Papineni et~al.(2002)Papineni, Roukos, Ward, and Zhu]{papineni2002bleu}
Kishore Papineni, Salim Roukos, Todd Ward, and Wei-Jing Zhu.
\newblock Bleu: a method for automatic evaluation of machine translation.
\newblock In \emph{Proceedings of the 40th annual meeting of the Association for Computational Linguistics}, 2002.

\bibitem[Park et~al.()Park, Chan, Zhang, Chiu, Zoph, Cubuk, and Le]{specaug}
Daniel~S. Park, William Chan, Yu~Zhang, Chung{-}Cheng Chiu, Barret Zoph, Ekin~D. Cubuk, and Quoc~V. Le.
\newblock Specaugment: {A} simple data augmentation method for automatic speech recognition.
\newblock In \emph{Interspeech 2019, 20th Annual Conference of the International Speech Communication Association, Graz, Austria, 15-19 September 2019}.

\bibitem[Peng et~al.(2023)Peng, Wang, Dong, Hao, Huang, Ma, and Wei]{kosmos2}
Zhiliang Peng, Wenhui Wang, Li~Dong, Yaru Hao, Shaohan Huang, Shuming Ma, and Furu Wei.
\newblock Kosmos-2: Grounding multimodal large language models to the world.
\newblock \emph{arXiv:2306.14824}, 2023.

\bibitem[Poria et~al.(2019)Poria, Hazarika, Majumder, Naik, Cambria, and Mihalcea]{Meld}
Soujanya Poria, Devamanyu Hazarika, Navonil Majumder, Gautam Naik, Erik Cambria, and Rada Mihalcea.
\newblock {MELD:} {A} multimodal multi-party dataset for emotion recognition in conversations.
\newblock In \emph{Proceedings of the 57th Conference of the Association for Computational Linguistics, {ACL} 2019, Florence, Italy, July 28- August 2, 2019, Volume 1: Long Papers}. Association for Computational Linguistics, 2019.

\bibitem[Qwen(2023)]{qwen7b}
Qwen.
\newblock Introducing qwen-7b: Open foundation and human-aligned models (of the state-of-the-arts), 2023.
\newblock URL \url{https://github.com/QwenLM/Qwen-7B}.

\bibitem[Radford et~al.(2023)Radford, Kim, Xu, Brockman, McLeavey, and Sutskever]{Whisper}
Alec Radford, Jong~Wook Kim, Tao Xu, Greg Brockman, Christine McLeavey, and Ilya Sutskever.
\newblock Robust speech recognition via large-scale weak supervision.
\newblock In \emph{International Conference on Machine Learning, {ICML} 2023, 23-29 July 2023, Honolulu, Hawaii, {USA}}, 2023.

\bibitem[Raffel et~al.(2020)Raffel, Shazeer, Roberts, Lee, Narang, Matena, Zhou, Li, and Liu]{raffel2020exploring}
Colin Raffel, Noam Shazeer, Adam Roberts, Katherine Lee, Sharan Narang, Michael Matena, Yanqi Zhou, Wei Li, and Peter~J Liu.
\newblock Exploring the limits of transfer learning with a unified text-to-text transformer.
\newblock \emph{The Journal of Machine Learning Research}, 2020.

\bibitem[Rubenstein et~al.()Rubenstein, Asawaroengchai, Nguyen, Bapna, Borsos, de~Chaumont~Quitry, Chen, Badawy, Han, Kharitonov, Muckenhirn, Padfield, Qin, Rozenberg, Sainath, Schalkwyk, Sharifi, Ramanovich, Tagliasacchi, Tudor, Velimirovic, Vincent, Yu, Wang, Zayats, Zeghidour, Zhang, Zhang, Zilka, and Frank]{AudioPALM}
Paul~K. Rubenstein, Chulayuth Asawaroengchai, Duc~Dung Nguyen, Ankur Bapna, Zal{\'{a}}n Borsos, F{\'{e}}lix de~Chaumont~Quitry, Peter Chen, Dalia~El Badawy, Wei Han, Eugene Kharitonov, Hannah Muckenhirn, Dirk Padfield, James Qin, Danny Rozenberg, Tara~N. Sainath, Johan Schalkwyk, Matthew Sharifi, Michelle~Tadmor Ramanovich, Marco Tagliasacchi, Alexandru Tudor, Mihajlo Velimirovic, Damien Vincent, Jiahui Yu, Yongqiang Wang, Vicky Zayats, Neil Zeghidour, Yu~Zhang, Zhishuai Zhang, Lukas Zilka, and Christian~Havn{\o} Frank.
\newblock Audiopalm: {A} large language model that can speak and listen.
\newblock \emph{CoRR}.

\bibitem[Shen et~al.(2023)Shen, Song, Tan, Li, Lu, and Zhuang]{HuggingGPT}
Yongliang Shen, Kaitao Song, Xu~Tan, Dongsheng Li, Weiming Lu, and Yueting Zhuang.
\newblock Hugginggpt: Solving {AI} tasks with chatgpt and its friends in huggingface.
\newblock \emph{CoRR}, abs/2303.17580, 2023.

\bibitem[Shi et~al.(2023)Shi, Chen, Zhang, and Yan]{shi2023achieving}
Xian Shi, Yanni Chen, Shiliang Zhang, and Zhijie Yan.
\newblock Achieving timestamp prediction while recognizing with non-autoregressive end-to-end asr model.
\newblock In \emph{National Conference on Man-Machine Speech Communication}. Springer, 2023.

\bibitem[Shu et~al.(2023)Shu, Dong, Chen, Huang, Zhang, Shi, Xiang, and Shi]{shu2023llasm}
Yu~Shu, Siwei Dong, Guangyao Chen, Wenhao Huang, Ruihua Zhang, Daochen Shi, Qiqi Xiang, and Yemin Shi.
\newblock Llasm: Large language and speech model.
\newblock \emph{arXiv:2308.15930}, 2023.

\bibitem[Sun et~al.(2023)Sun, Yu, Cui, Zhang, Zhang, Wang, Gao, Liu, Huang, and Wang]{EMU}
Quan Sun, Qiying Yu, Yufeng Cui, Fan Zhang, Xiaosong Zhang, Yueze Wang, Hongcheng Gao, Jingjing Liu, Tiejun Huang, and Xinlong Wang.
\newblock Generative pretraining in multimodality.
\newblock \emph{arXiv:2307.05222}, 2023.

\bibitem[Touvron et~al.(2023{\natexlab{a}})Touvron, Lavril, Izacard, Martinet, Lachaux, Lacroix, Rozi{\`e}re, Goyal, Hambro, Azhar, et~al.]{llama}
Hugo Touvron, Thibaut Lavril, Gautier Izacard, Xavier Martinet, Marie-Anne Lachaux, Timoth{\'e}e Lacroix, Baptiste Rozi{\`e}re, Naman Goyal, Eric Hambro, Faisal Azhar, et~al.
\newblock {LLaMA}: Open and efficient foundation language models.
\newblock \emph{arXiv:2302.13971}, 2023{\natexlab{a}}.

\bibitem[Touvron et~al.(2023{\natexlab{b}})Touvron, Lavril, Izacard, Martinet, Lachaux, Lacroix, Rozi{\`e}re, Goyal, Hambro, Azhar, et~al.]{touvron2023llama}
Hugo Touvron, Thibaut Lavril, Gautier Izacard, Xavier Martinet, Marie-Anne Lachaux, Timoth{\'e}e Lacroix, Baptiste Rozi{\`e}re, Naman Goyal, Eric Hambro, Faisal Azhar, et~al.
\newblock Llama: Open and efficient foundation language models.
\newblock \emph{arXiv:2302.13971}, 2023{\natexlab{b}}.

\bibitem[Touvron et~al.(2023{\natexlab{c}})Touvron, Martin, Stone, Albert, Almahairi, Babaei, Bashlykov, Batra, Bhargava, Bhosale, Bikel, Blecher, Canton{-}Ferrer, Chen, Cucurull, Esiobu, Fernandes, Fu, Fu, Fuller, Gao, Goswami, Goyal, Hartshorn, Hosseini, Hou, Inan, Kardas, Kerkez, Khabsa, Kloumann, Korenev, Koura, Lachaux, Lavril, Lee, Liskovich, Lu, Mao, Martinet, Mihaylov, Mishra, Molybog, Nie, Poulton, Reizenstein, Rungta, Saladi, Schelten, Silva, Smith, Subramanian, Tan, Tang, Taylor, Williams, Kuan, Xu, Yan, Zarov, Zhang, Fan, Kambadur, Narang, Rodriguez, Stojnic, Edunov, and Scialom]{llama2}
Hugo Touvron, Louis Martin, Kevin Stone, Peter Albert, Amjad Almahairi, Yasmine Babaei, Nikolay Bashlykov, Soumya Batra, Prajjwal Bhargava, Shruti Bhosale, Dan Bikel, Lukas Blecher, Cristian Canton{-}Ferrer, Moya Chen, Guillem Cucurull, David Esiobu, Jude Fernandes, Jeremy Fu, Wenyin Fu, Brian Fuller, Cynthia Gao, Vedanuj Goswami, Naman Goyal, Anthony Hartshorn, Saghar Hosseini, Rui Hou, Hakan Inan, Marcin Kardas, Viktor Kerkez, Madian Khabsa, Isabel Kloumann, Artem Korenev, Punit~Singh Koura, Marie{-}Anne Lachaux, Thibaut Lavril, Jenya Lee, Diana Liskovich, Yinghai Lu, Yuning Mao, Xavier Martinet, Todor Mihaylov, Pushkar Mishra, Igor Molybog, Yixin Nie, Andrew Poulton, Jeremy Reizenstein, Rashi Rungta, Kalyan Saladi, Alan Schelten, Ruan Silva, Eric~Michael Smith, Ranjan Subramanian, Xiaoqing~Ellen Tan, Binh Tang, Ross Taylor, Adina Williams, Jian~Xiang Kuan, Puxin Xu, Zheng Yan, Iliyan Zarov, Yuchen Zhang, Angela Fan, Melanie Kambadur, Sharan Narang, Aur{\'{e}}lien Rodriguez, Robert Stojnic, Sergey Edunov,
  and Thomas Scialom.
\newblock Llama 2: Open foundation and fine-tuned chat models.
\newblock \emph{CoRR}, abs/2307.09288, 2023{\natexlab{c}}.

\bibitem[Vaswani et~al.(2017)Vaswani, Shazeer, Parmar, Uszkoreit, Jones, Gomez, Kaiser, and Polosukhin]{Transformer}
Ashish Vaswani, Noam Shazeer, Niki Parmar, Jakob Uszkoreit, Llion Jones, Aidan~N. Gomez, Lukasz Kaiser, and Illia Polosukhin.
\newblock Attention is all you need.
\newblock In Isabelle Guyon, Ulrike von Luxburg, Samy Bengio, Hanna~M. Wallach, Rob Fergus, S.~V.~N. Vishwanathan, and Roman Garnett, editors, \emph{Advances in Neural Information Processing Systems 30: Annual Conference on Neural Information Processing Systems 2017}, 2017.

\bibitem[Vedantam et~al.(2015)Vedantam, Lawrence~Zitnick, and Parikh]{vedantam2015cider}
Ramakrishna Vedantam, C~Lawrence~Zitnick, and Devi Parikh.
\newblock Cider: Consensus-based image description evaluation.
\newblock In \emph{CVPR}, 2015.

\bibitem[Wang et~al.(2020)Wang, Wu, and Pino]{CoVoST2}
Changhan Wang, Anne Wu, and Juan~Miguel Pino.
\newblock Covost 2: {A} massively multilingual speech-to-text translation corpus.
\newblock abs/2007.10310, 2020.
\newblock URL \url{https://arxiv.org/abs/2007.10310}.

\bibitem[Wang et~al.(2023{\natexlab{a}})Wang, Liao, Huang, Lu, Wu, Liu, Zong, and Zhang]{wang2023blsp}
Chen Wang, Minpeng Liao, Zhongqiang Huang, Jinliang Lu, Junhong Wu, Yuchen Liu, Chengqing Zong, and Jiajun Zhang.
\newblock Blsp: Bootstrapping language-speech pre-training via behavior alignment of continuation writing.
\newblock \emph{arXiv:2309.00916}, 2023{\natexlab{a}}.

\bibitem[Wang et~al.(2023{\natexlab{b}})Wang, Han, Shafran, Wu, Chiu, Cao, Wang, Chen, Zhang, Soltau, Rubenstein, Zilka, Yu, Meng, Pundak, Siddhartha, Schalkwyk, and Wu]{SLM}
Mingqiu Wang, Wei Han, Izhak Shafran, Zelin Wu, Chung{-}Cheng Chiu, Yuan Cao, Yongqiang Wang, Nanxin Chen, Yu~Zhang, Hagen Soltau, Paul~K. Rubenstein, Lukas Zilka, Dian Yu, Zhong Meng, Golan Pundak, Nikhil Siddhartha, Johan Schalkwyk, and Yonghui Wu.
\newblock {SLM:} bridge the thin gap between speech and text foundation models.
\newblock abs/2310.00230, 2023{\natexlab{b}}.

\bibitem[Wang et~al.(2023{\natexlab{c}})Wang, Han, Shafran, Wu, Chiu, Cao, Wang, Chen, Zhang, Soltau, et~al.]{wang2023slm}
Mingqiu Wang, Wei Han, Izhak Shafran, Zelin Wu, Chung-Cheng Chiu, Yuan Cao, Yongqiang Wang, Nanxin Chen, Yu~Zhang, Hagen Soltau, et~al.
\newblock Slm: Bridge the thin gap between speech and text foundation models.
\newblock \emph{arXiv:2310.00230}, 2023{\natexlab{c}}.

\bibitem[Wang et~al.(2023{\natexlab{d}})Wang, Zhou, Zhang, Wu, Liu, Gaur, Chen, Li, and Wei]{VIOLA}
Tianrui Wang, Long Zhou, Ziqiang Zhang, Yu~Wu, Shujie Liu, Yashesh Gaur, Zhuo Chen, Jinyu Li, and Furu Wei.
\newblock Viola: Unified codec language models for speech recognition, synthesis, and translation.
\newblock \emph{CoRR}, 2023{\natexlab{d}}.

\bibitem[Wang et~al.(2023{\natexlab{e}})Wang, Thakker, Chen, Kanda, Eskimez, Chen, Tang, Liu, Li, and Yoshioka]{SpeechX}
Xiaofei Wang, Manthan Thakker, Zhuo Chen, Naoyuki Kanda, Sefik~Emre Eskimez, Sanyuan Chen, Min Tang, Shujie Liu, Jinyu Li, and Takuya Yoshioka.
\newblock Speechx: Neural codec language model as a versatile speech transformer.
\newblock \emph{CoRR}, 2023{\natexlab{e}}.

\bibitem[Wu et~al.(2023{\natexlab{a}})Wu, Gaur, Chen, Zhou, Zhu, Wang, Li, Liu, Ren, Liu, and Wu]{speechllama}
Jian Wu, Yashesh Gaur, Zhuo Chen, Long Zhou, Yimeng Zhu, Tianrui Wang, Jinyu Li, Shujie Liu, Bo~Ren, Linquan Liu, and Yu~Wu.
\newblock On decoder-only architecture for speech-to-text and large language model integration.
\newblock abs/2307.03917, 2023{\natexlab{a}}.

\bibitem[Wu et~al.(2023{\natexlab{b}})Wu, Fei, Qu, Ji, and Chua]{NextGPT}
Shengqiong Wu, Hao Fei, Leigang Qu, Wei Ji, and Tat{-}Seng Chua.
\newblock Next-gpt: Any-to-any multimodal {LLM}.
\newblock \emph{CoRR}, abs/2309.05519, 2023{\natexlab{b}}.

\bibitem[Zeghidour et~al.(2022)Zeghidour, Luebs, Omran, Skoglund, and Tagliasacchi]{SoundStream}
Neil Zeghidour, Alejandro Luebs, Ahmed Omran, Jan Skoglund, and Marco Tagliasacchi.
\newblock Soundstream: An end-to-end neural audio codec.
\newblock \emph{{IEEE} {ACM} Trans. Audio Speech Lang. Process.}, 2022.

\bibitem[Zhang et~al.(2023{\natexlab{a}})Zhang, Li, Zhang, Zhan, Wang, Zhou, and Qiu]{speechgpt}
Dong Zhang, Shimin Li, Xin Zhang, Jun Zhan, Pengyu Wang, Yaqian Zhou, and Xipeng Qiu.
\newblock Speechgpt: Empowering large language models with intrinsic cross-modal conversational abilities.
\newblock \emph{CoRR}, abs/2305.11000, 2023{\natexlab{a}}.

\bibitem[Zhang et~al.(2023{\natexlab{b}})Zhang, Zhang, Li, Zhou, and Qiu]{SpeechTokenizer}
Xin Zhang, Dong Zhang, Shimin Li, Yaqian Zhou, and Xipeng Qiu.
\newblock Speechtokenizer: Unified speech tokenizer for speech large language models.
\newblock \emph{CoRR}, abs/2308.16692, 2023{\natexlab{b}}.

\bibitem[Zhang et~al.(2023{\natexlab{c}})Zhang, Han, Qin, Wang, Bapna, Chen, Chen, Li, Axelrod, Wang, Meng, Hu, Rosenberg, Prabhavalkar, Park, Haghani, Riesa, Perng, Soltau, Strohman, Ramabhadran, Sainath, Moreno, Chiu, Schalkwyk, Beaufays, and Wu]{USM}
Yu~Zhang, Wei Han, James Qin, Yongqiang Wang, Ankur Bapna, Zhehuai Chen, Nanxin Chen, Bo~Li, Vera Axelrod, Gary Wang, Zhong Meng, Ke~Hu, Andrew Rosenberg, Rohit Prabhavalkar, Daniel~S. Park, Parisa Haghani, Jason Riesa, Ginger Perng, Hagen Soltau, Trevor Strohman, Bhuvana Ramabhadran, Tara~N. Sainath, Pedro~J. Moreno, Chung{-}Cheng Chiu, Johan Schalkwyk, Fran{\c{c}}oise Beaufays, and Yonghui Wu.
\newblock Google usm: Scaling automatic speech recognition beyond 100 languages.
\newblock \emph{CoRR}, 2023{\natexlab{c}}.

\bibitem[Zhou et~al.(2022)Zhou, Wang, Cui, Zhang, Yan, Zhou, and Zhou]{mmspeech}
Xiaohuan Zhou, Jiaming Wang, Zeyu Cui, Shiliang Zhang, Zhijie Yan, Jingren Zhou, and Chang Zhou.
\newblock Mmspeech: Multi-modal multi-task encoder-decoder pre-training for speech recognition.
\newblock abs/2212.00500, 2022.

\end{thebibliography}

%%%%%%%%%%%%%%%%%%%%%%%%%%%%%%%%%%%%%%%%%%%%%%%%%%%%%%%%%%%%

\clearpage
\newpage

\appendix

\section{Hyperparameters}
\label{appendix:hyperparam}
We report the detailed training hyperparameter settings of Qwen-Audio in Table~\ref{tab:hyperparam}.

\begin{table}[htbp]
    \centering
    \caption{Training hyperparameters of Qwen-Audio}
    \begin{tabular}{l cc}
         \toprule
         Configuration            &  Multi-task Pre-training & Supervised Fine-tuning \\
         \midrule
         Audio encoder init.      & Whisper-large-v2 & Qwen-audio 1st-stage \\
         LLM init.                & Qwen-7B & Qwen-7B \\
         SpecAugment Policy       & LibriSpeech Basic & LibriSpeech Basic \\
         Optimizer                & AdamW & AdamW \\
         Optimizer hyperparameter & \multicolumn{2}{c}{$\beta_{1}$=0.9, $\beta_{2}$=0.98, $eps=1e^{-6}$} \\
         Peak learning rate       & $5e^{-5}$ & $1e^{-5}$ \\
         Minimum learning rate    & $1e^{-5}$ & $1e^{-6}$ \\
         Audio encoder learning rate decay  & 0.95 & 0 \\
         Learning rate schedule   & cosine decay & cosine decay \\
         Weight decay             & 0.05 & 0.05 \\
         Gradient clip            & 1.0 & 1.0 \\
         Training steps           & 500k & 8k \\
         Warm-up steps            & 2000 & 3k \\
         Global batch size        & 120 & 128 \\
         Gradient Acc.            & 1 & 8 \\
         Numerical precision      & {$\mathtt{bfloat16}$} & {$\mathtt{bfloat16}$}\\
         Optimizer sharding       & {\ding{51}} & {\ding{51}}\\
         Activation checkpointing & {\ding{55}} & {\ding{55}}\\
         Model parallelism        & \ding{55} & 2 \\
         Pipeline parallelism     & {\ding{55}}  & {\ding{55}}\\
         \bottomrule
    \end{tabular}
    \label{tab:hyperparam}
\end{table}

\end{document}